\documentclass[runningheads]{llncs}

\usepackage[T1]{fontenc}
\usepackage[protrusion=false,expansion=true]{microtype}
\usepackage{graphicx}
\usepackage{amsmath,amssymb}
\usepackage{cite}
\usepackage{booktabs}
\usepackage{multirow}
\usepackage{makecell}
\usepackage{needspace}
\usepackage{enumitem}
\newenvironment{packeditemize}
  {\begin{itemize}[leftmargin=*,itemsep=2pt,parsep=0pt,topsep=3pt,partopsep=0pt]}
  {\end{itemize}}
\usepackage{threeparttable}
\usepackage[table]{xcolor}
\usepackage{pifont}
\usepackage{tikz}
\usepackage{algorithm}
\usepackage{wrapfig}
\usepackage{algpseudocode}
\usepackage[hidelinks]{hyperref}
\usepackage{xurl}

\makeatletter
\renewcommand\subsubsection{\@startsection{subsubsection}{3}{\z@}%
  {-3pt}{-0.5em}{\normalfont\normalsize\bfseries\boldmath}}
\renewcommand\paragraph{\@startsection{paragraph}{4}{\z@}%
  {-3pt}{-0.5em}{\normalfont\normalsize\itshape}}
\makeatother

\usetikzlibrary{arrows.meta,positioning,fit,calc,backgrounds}
\usetikzlibrary{fadings,shadings}

\definecolor{agentorange}{RGB}{242,140,35}
\definecolor{agentgold}{RGB}{232,166,14}
\definecolor{agentrose}{RGB}{219,79,147}
\definecolor{agentviolet}{RGB}{131,56,236}

\pgfdeclarehorizontalshading{agentrainbow}{100bp}{%
  color(0bp)=(agentorange);
  color(25bp)=(agentorange);
  color(28bp)=(agentorange);
  color(31bp)=(agentrose);
  color(34bp)=(agentviolet);
  color(55bp)=(agentviolet);
  color(62.5bp)=(agentrose);
  color(69bp)=(agentorange);
  color(75bp)=(agentgold);
  color(100bp)=(agentgold)%
}

\newcounter{agentbankrainbow}

\definecolor{flowpurple}{HTML}{8767A5}
\DeclareRobustCommand{\flowstep}[1]{%
  \tikz[baseline=(flowlabel.base)]{%
    \node[circle,fill=flowpurple,text=white,
      inner sep=0pt,outer sep=0pt,minimum size=11pt,
      font=\sffamily\bfseries\scriptsize,
      text height=1.4ex,text depth=0ex] (flowlabel) {#1};%
  }\hspace{0.35em}%
}
\definecolor{tableblue}{HTML}{205B78}
\definecolor{tablefill}{HTML}{E6F0F5}
\definecolor{tablegreen}{HTML}{009900}
\definecolor{tablered}{HTML}{D90000}
\definecolor{tableorange}{HTML}{F07800}
\newcommand{\yes}{\textcolor{tablegreen}{\ding{51}}}
\newcommand{\no}{\textcolor{tablered}{\ding{55}}}
\newcommand{\partl}{\textcolor{tableorange}{$\circ$}}
\newcommand{\Req}{\textbf{require}\ }

\usepackage{mathrsfs}

\begin{document}

\title{AgentBank: Banking on Blockchain Agentic Payments}
\titlerunning{AgentBank: Banking on Blockchain Agentic Payments}


\authorrunning{Xiang et al.}

\author{
Yuexin Xiang\inst{1,2} \and
Martin Pere\v{s}\'{i}ni\inst{3} \and
Yuchen Lei\inst{4} \and
Yuquan Li\inst{5,2} \and \\
Yuanzhe Zhang\inst{6} \and
Andreas Deppeler\inst{7} \and
Tsz Hon Yuen\inst{1} \and
Shiping Chen\inst{8} \and \\
Qin Wang\inst{2,8} \and
Burkhard Stiller\inst{3}
}

\institute{
Faculty of Information Technology, Monash University, Australia
\and
Digital Finance CRC, Australia
\and
Communication Systems Group, Department of Informatics,
University of Zurich, Switzerland
\and
School of Cyber Science and Engineering,
Wuhan University, China
\and
Faculty of Business and Economics,
The University of Melbourne, Australia
\and
Digital Trust Centre, Nanyang Technological University,
Singapore
\and
School of Business, Monash University, Malaysia
\and
CSIRO, Sydney, Australia
}

\maketitle

\begin{abstract}

AI agents increasingly pay for data and computation through protocols such as x402 and AP2, but individually funded wallets idle and require asset transfers for every payment. We present \textit{AgentBank}, a smart-contract banking layer that lets agents spend bank liabilities under owner-issued mandates and on-chain spending limits. Internal payments update account balances without transferring assets, while pooled backing funds are allocated between reserves and yield-bearing strategies.
We bound losses from compromised agent keys and analyze how correlated demand affects reserve requirements. Experiments show that AgentBank reduces idle liquidity by 52--77\% relative to per-agent wallets and lowers gas costs for batched internal payments. Using historical DeFi liquidity data, we evaluate how reserve policies balance treasury returns against payment availability during market stress. We further examine the risks of automated bank runs and evaluate mechanisms to mitigate them.



\end{abstract}

\section{Introduction}
\label{sec:intro}

AI agents increasingly purchase API calls, data, inference, and the services of other agents with limited human involvement, and a growing set of protocols lets them authorize payments: x402 revives the HTTP~402 status code for per-request stablecoin payments~\cite{x402foundation2025whitepaper}, AP2 conveys user mandates to merchants~\cite{google2025ap2}, and card networks and payment processors have released comparable frameworks~\cite{visa2025trustedAgentProtocol,weinstein2026mpp}. On Base, x402 recorded 136.7~million settlements in 280~days, although most of this count is fictitious or operator-internal rather than independent demand~\cite{ling2026agenticCommerceMeasurement}.

These systems share one financial abstraction, the \emph{wallet}. An agent's payment capacity is the balance of a wallet funded for that agent, possibly guarded by spending policies~\cite{coinbase2026agenticWallets}. 
This abstraction creates three costs.
First, liquidity is fragmented.
Each wallet must be prefunded for its own peak demand, and task-driven agent demand is bursty, so the sum of per-agent buffers far exceeds what aggregate demand requires. Second, prefunded balances are idle and earn nothing. Third, every payment, even between agents of the same ecosystem, transfers the settlement asset on-chain. 
In addition, a security concern arises that an agent manipulated through prompt or memory injection~\cite{patlan2026fakememories} can spend whatever its wallet holds unless limits are enforced below the model.

Banking resolves the first three costs by construction. A bank balance is a liability rather than ownership of specific assets.
Payments between accounts reassign liabilities without moving reserves, and the bank holds liquid reserves only against net outflows, deploying the remainder productively~\cite{diamond1983bankRuns,kashyap2002liquidityProviders}. Tokenized deposits carry this separation onto programmable ledgers~\cite{garratt2023tokenisedDeposits,bis2023blueprint,bis2026agora,jpmorgan2025jpmd}. The same transformation creates fragility: backing assets deployed into DeFi venues can become illiquid when they are needed most. After the KelpDAO exploit of 18~April~2026, the USDC market of Aave~v3 on Ethereum remained at 99.87\% utilization for four days, leaving suppliers of \$1.89~billion unable to withdraw~\cite{liao2026arfc,galaxy2026kelp}. Fractional reserves also invite runs~\cite{diamond1983bankRuns,goldstein2005demanddeposit}, and on-chain transparency is known to accelerate them~\cite{liu2023anatomy,ahmed2024stablecoinRuns}, a concern that sharpens when depositors are themselves automated agents.

Deployed agent-payment systems occupy two corners of a design space: a funded wallet holds a full reserve for each agent, and an agent that holds yield-bearing tokens invests its entire balance. This work presents \textbf{AgentBank}
, a smart-contract banking layer for AI agents that occupies the interior, where pooled liabilities are backed by a reserve whose size is a policy choice, and studies the efficiency--resilience trade-off that this choice introduces. The problem joins stablecoin reserve design, delegated strategy management, and transaction ordering in a setting where both payers and depositors are machines. \mbox{We make three contributions.}
\begin{enumerate}[label=\textbf{\arabic*.)}]
\item We design AgentBank, a smart-contract banking architecture that allows agents to spend bank liabilities under owner-issued mandates with on-chain spending limits. Payments between accounts are settled internally, while pooled backing assets are allocated between a settlement reserve and approved yield-bearing strategies under on-chain reserve and haircut-adjusted coverage constraints. Uncovered external payments enter an expiring queue. Solvency-contingent redemption governs payouts after impairment. Extensions include forward pricing, epoch settlement, and liquidity fees.

\item We prove balance-sheet conservation and bound losses from compromised agent keys. We show that pooled reserve requirements converge to a single-factor (Vasicek) limit under correlated demand, derive a granularity adjustment for finite pools, and establish why the conventional square-root law underestimates liquidity requirements. We further show that sequential redemption at par concentrates losses on depositors who redeem later, while solvency-contingent redemption eliminates this first-mover advantage after loss recognition, absent further losses or liquidation costs.

\item We evaluate a Solidity prototype against EIP-3009 and ERC-4337 baselines to measure the gas cost of delegated payments. Using bursty, correlated agent demand and historical lending-pool data, we quantify idle-liquidity savings of 52--77\% relative to per-agent wallets and attribute them to owner pooling, bank pooling, and payment internalization. We also characterize the trade-off between treasury returns and payment availability during the April~2026 liquidity freeze, estimate reserve requirements for Treasury-bill-only strategies, and measure how fast automated depositors exhaust reserves.
\end{enumerate}

\section{Background and Related Work}
\label{sec:related}

\begin{wraptable}{r}{0.50\textwidth}
\centering
\begin{threeparttable}
\caption{Design comparison.}
\label{tab:comparison}
\scriptsize
\setlength{\tabcolsep}{1.0pt}
\renewcommand{\arraystretch}{1.12}
\arrayrulecolor{black}
\newcommand{\comparisonhead}[1]{\rotatebox[origin=l]{90}{\textbf{#1}}}
\begin{tabular*}{\linewidth}{@{\extracolsep{\fill}}lcccccc@{}}
\toprule
\textbf{Design} & \comparisonhead{Par} & \comparisonhead{No prefunding} & \comparisonhead{Internal pay.} & \comparisonhead{Yield} & \comparisonhead{On-chain caps} & \comparisonhead{Illiquidity} \\
\midrule
\makecell[l]{Funded agent\\wallet~\cite{x402foundation2025whitepaper}} & \yes & \no & \no & \no & \partl & n/a (reserved) \\
\makecell[l]{Smart account,\\session key} & \yes & \partl & \no & \no & \yes & n/a (reserved) \\
\makecell[l]{Channels,\\sessions~\cite{weinstein2026mpp}} & \yes & \no & \partl & \no & \partl & n/a (per pair) \\
TEE netting~\cite{circle2026nanopayments} & \yes & \partl & \yes & \no & \partl & n/a (reserved) \\
Yield-bearing tokens & \no & \no & \yes & \yes & \no & blocked \\
\makecell[l]{Off-chain custodial\\ledger} & \yes & \yes & \yes & \partl & \no & opaque \\
\midrule
\textbf{AgentBank} & \yes & \yes & \yes & \yes & \yes & reserve, queue \\
\bottomrule
\end{tabular*}
\begin{tablenotes}[flushleft]
\scriptsize
\item \yes: provided. \partl: partially provided.\newline \no: not provided. N/a: not applicable.
\item Prefunding is per agent. Smart accounts share it per owner.
\end{tablenotes}
\end{threeparttable}
\end{wraptable}

\subsubsection{Agent Payments.}
A systematization of agent-to-agent payments on blockchains decomposes the payment lifecycle into four stages and identifies weak intent binding as an open problem~\cite{zhang2026sokA2APayments}. Its taxonomy contains no category for pooled, bank-like liabilities, which is the design space of this work. Deployed protocols bind payments to HTTP requests (x402, settled by EIP-3009 transfer authorizations~\cite{x402foundation2025whitepaper}), to verifiable user mandates (AP2~\cite{google2025ap2}), or to card credentials~\cite{visa2025trustedAgentProtocol}. Security studies identify risks in x402 payments and context-manipulation attacks on agents~\cite{wang2026x402risks,patlan2026fakememories}. Account-abstraction and delegation standards make accounts programmable, scope delegations and recurring allowances, and register agent identities. In all of these systems, the funds behind an authorization are the balance of the authorized wallet or account. AgentBank instead attaches delegated authority to a \emph{bank liability} whose backing is managed elsewhere.

\subsubsection{Settlement Internalization.}
Payment channels defer base-layer settlement~\cite{miller2019sprites,dziembowski2019perun,sivaraman2020spider} but lock liquidity per channel and require rebalancing~\cite{khalil2017revive}. Agent-oriented designs batch or net settlement: Circle Gateway nets signed authorizations in a trusted execution environment~\cite{circle2026nanopayments}, and A402 aggregates channel settlement through a TEE-based vault~\cite{li2026a402}. These designs keep payer funds fully reserved and idle. In conventional payment systems, tiering and liquidity-saving mechanisms reduce liquidity needs~\cite{kahn2009tiering,martin2008liquiditySaving,johnson2004economizing}, and simulations of tiering in CHAPS attribute most of the saving to pooling rather than internalization~\cite{lasaosa2008tiering}, as our decomposition does for agent demand. The underlying effect is risk pooling, whose classic form is the square-root law of inventory pooling~\cite{eppen1979centralization}. Pooling gains shrink under dependent or heavy-tailed demand~\cite{corbett2006pooling,bimpikis2016heavytailed}.

\subsubsection{Programmable Banking.}
Tokenized deposits preserve the distinction between account liabilities and backing assets on programmable ledgers~\cite{garratt2023tokenisedDeposits,bis2023blueprint,bis2025nextgen,bis2026agora}, and a regulated bank already issues a deposit token on a public chain~\cite{jpmorgan2025jpmd}. Stablecoins behave like money-market funds under stress~\cite{anadu2023runs}, runs are accelerated by on-chain transparency~\cite{liu2023anatomy}, the effect of disclosure on run risk is ambiguous~\cite{ahmed2024stablecoinRuns}, redemption access trades price stability against run risk~\cite{ma2025stablecoinRuns}, and capital and liquidity thresholds are the main regulatory levers~\cite{goel2026stablecoinRegulation}. DeFi lending pools make deposits productive but limit withdrawals to unborrowed liquidity~\cite{gudgeon2020loanable,werner2022sok}, which can vanish in self-reinforcing episodes~\cite{chiu2023fragility,lehar2022fragility}. ERC-4626 vaults and their asynchronous extension ERC-7540 standardize share accounting and queued redemptions. Vault shares, however, float in value and pass strategy losses to holders. \autoref{tab:comparison} contrasts these designs with AgentBank.

\Needspace{12\baselineskip}
\section{AgentBank Design}
\label{sec:design}

We build AgentBank from two core smart contracts, BankCore and TreasuryManager, plus an adapter for each approved investment strategy. BankCore handles accounts and payments. TreasuryManager manages their pooled backing assets through the adapters. We describe their components first, then explain how they implement the bank's operations.

\suppressfloats[t]
\begin{figure}[t]
\centering
\includegraphics[width=\linewidth]{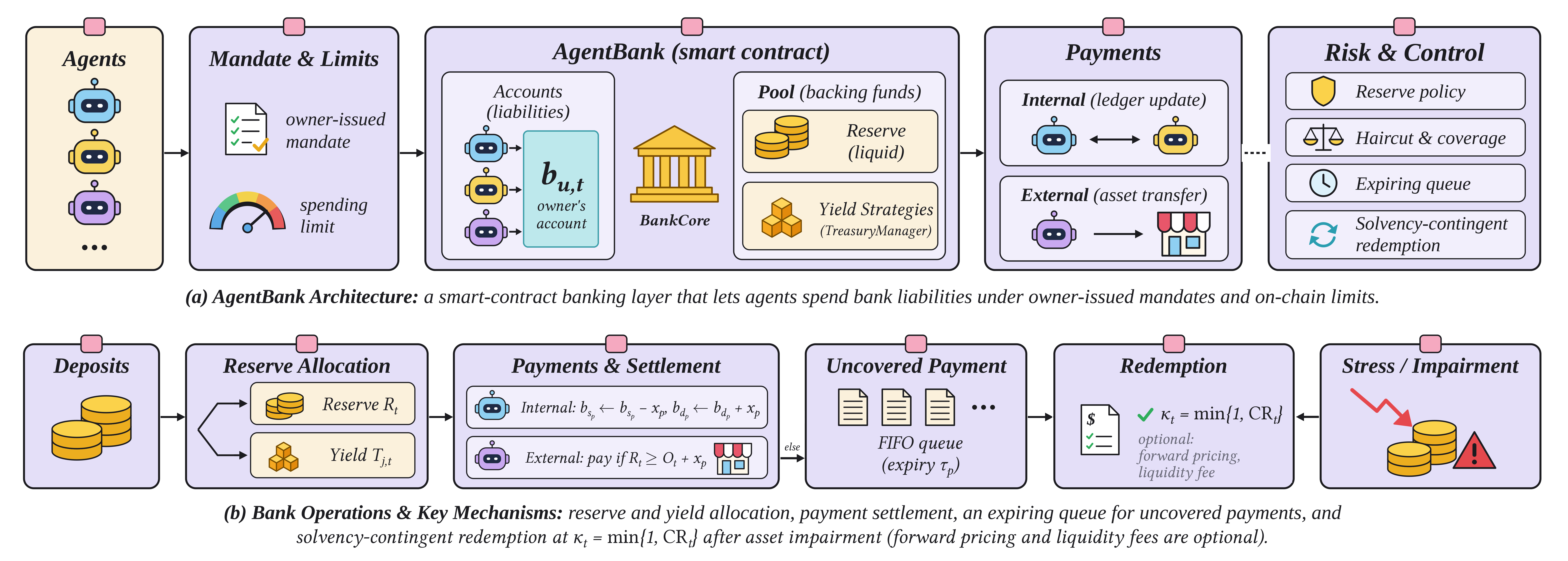}
\caption{AgentBank: \textbf{(a)}~contract components; \textbf{(b)}~operational flow.\newline
(\mbox{\flowstep{1}deposits and allocation},
\mbox{\flowstep{2}payments},
\mbox{\flowstep{3}queueing},
\mbox{\flowstep{4}impairment}.)}
\label{fig:arch}
\end{figure}

\subsection{Contract Components (\autoref{fig:arch}(a))}
\label{sec:roles}
\label{sec:balance}

Each settlement asset has its own BankCore and TreasuryManager. With one approved strategy, the deployment contains three AgentBank contracts:
\begin{packeditemize}
\item[$\bullet$] \textbf{BankCore.} Accepts deposits, stores owner balances and mandates, checks signed payment intents, and holds the liquid reserve. It implements the Accounts, Mandate \& Limits, and Payments blocks, including internal transfers, external payouts, the FIFO queue, and impairment recovery.
\item[$\bullet$] \textbf{TreasuryManager.} Moves backing assets between BankCore and approved strategies. It implements Reserve Allocation and manages Yield Strategies, checking the reserve floor and haircut-adjusted coverage before releasing funds.
\item[$\bullet$] \textbf{Strategy adapter.} Sits behind Yield Strategies and connects TreasuryManager to one external venue. The prototype's ERC-4626 adapter invests and withdraws assets and reports position values and withdrawable amounts.
\end{packeditemize}
The Risk \& Control block groups BankCore's settlement rules with TreasuryManager's deployment constraints.

Account owners deposit assets and authorize agents to spend from their accounts. Several agents may share one owner's balance. Relayers submit signed intents, and external recipients receive asset transfers. The bank operator supplies first-loss capital, approves strategies, and sets treasury parameters, but cannot modify account balances.

We assume that the published contracts execute correctly and that the settlement asset retains its nominal value. Depeg and issuer freezes are outside the model. Adversaries may compromise an agent's session key or manipulate the agent into signing harmful intents~\cite{patlan2026fakememories}. They may censor, delay, or reorder transactions, as relayers and block builders can~\cite{daian2020flashboys}, and replay intents across mandates, deployments, or chains. They may also cause strategy losses or illiquidity, including through exploits~\cite{gervais2026a1}. Operator misconduct is limited to parameter and strategy choices within the on-chain treasury constraints. Parameter changes should be subject to a timelock.

\subsubsection{Accounts and Backing Assets.}
The Accounts and Pool blocks describe the two sides of the bank's balance sheet. BankCore records account liabilities and holds the reserve. The adapters hold invested positions and report their values through TreasuryManager. Each deployment uses one ERC-20 settlement asset, typically a fiat-denominated stablecoin, and performs no conversion between assets. We fix one asset $\sigma$ and let index $t$ denote the contract state immediately before an operation. BankCore maintains a set of accounts $\mathcal{U}$, where $b_{u,t}\geq0$ denotes the balance of account $u\in\mathcal{U}$. This balance is a liability of AgentBank rather than a claim on a dedicated reserve. External payments accepted but not yet settled form outstanding obligations $O_t$, held in a first-in, first-out (FIFO) queue. Liabilities are
\begin{equation}
D_t=\textstyle\sum_{u\in\mathcal{U}}b_{u,t},\qquad L_t=D_t+O_t .
\end{equation}
On the asset side, $R_t$ is the settlement reserve held by BankCore, and $T_{j,t}$ is the marked value of the position in approved strategy $j\in\{1,\dots,J\}$. Total assets and equity are
\begin{equation}
A_t=R_t+\textstyle\sum_{j}T_{j,t},\qquad E_t=A_t-L_t ,
\end{equation}
where the operator supplies initial equity $E_0>0$ as a first-loss buffer. Each strategy carries a valuation factor $h_j\in[0,1]$ (with $h_j=1$ meaning no haircut), giving haircut-adjusted effective assets and the coverage ratio
\begin{equation}
\begin{aligned}
A^{\mathrm{eff}}_t &= R_t+\textstyle\sum_j h_jT_{j,t}, \\
\mathrm{CR}_t &= A^{\mathrm{eff}}_t/L_t \qquad (L_t>0).
\end{aligned}
\label{eq:cr}
\end{equation}

\subsection{Operational Flow (\autoref{fig:arch}(b))}
\label{sec:operations}
\label{sec:treasury}
\label{sec:mandate}
\label{sec:boundary}
\label{sec:impairment}

The flow runs from deposits and allocation to authorized payments. Queue processing handles insufficient liquidity, while impairment handling adjusts both immediate and queued payouts after asset losses.

\subsubsection{\flowstep{1}Deposits and Allocation.}
An owner deposits the settlement asset into BankCore, which increases the owner's account balance and its reserve by the same amount. In the Reserve Allocation block, the operator separately triggers treasury operations, choosing when and how much to invest or recall. For a deployment of $z>0$ units into strategy $j$, TreasuryManager allows BankCore to transfer those assets to the adapter only if
\begin{equation}
\begin{aligned}
R_t-z &\geq O_t+r_{\min}D_t, \\
A^{\mathrm{eff}}_t-(1-h_j)\,z &\geq \mathrm{CR}_{\min}L_t.
\end{aligned}
\label{eq:deploy}
\end{equation}
where $r_{\min}\in[0,1]$ is the minimum reserve ratio and $\mathrm{CR}_{\min}\geq 1$ the coverage floor. The adapter then invests the released assets. To recall funds, TreasuryManager checks the adapter's withdrawable amount and instructs it to return assets to BankCore, which updates its reserve. Within these constraints, the operator targets a reserve $R^*_t=O_t+B_t$, where $B_t$ is the buffer for new external demand. The \emph{fixed} policy sets $B_t=r_{\mathrm{base}}D_t$ with $r_{\mathrm{base}}\in[r_{\min},1]$. The \emph{demand-aware} policy sets
\begin{equation}
B_t=\max\{r_{\min}D_t,\;Q_\alpha(\mathcal{X}_{t,W})\},
\label{eq:demand_aware}
\end{equation}
where $\mathcal{X}_{t,W}$ contains rolling $\Delta$-hour totals of requested external outflows (payments and withdrawals) observed over the preceding $W$ days. Thus $\Delta$ is the planning horizon in hours, $W$ is the look-back duration in days, and $Q_\alpha$ is the empirical $\alpha$-quantile. The demand estimate is computed off-chain. The deployment constraints in \autoref{eq:deploy} remain enforced on-chain.

Coverage does not imply liquidity: the reserve and assets that can be recalled in time must cover near-term outflows, as in the Basel liquidity coverage ratio~\cite{bcbs2013lcr}. Lending-pool recalls depend on unborrowed liquidity~\cite{gudgeon2020loanable}, which can approach zero within hours. Haircuts should reflect stressed withdrawability and price risk.

\subsubsection{\flowstep{2}Payment Execution.}
Payment $p$ draws from the mandate's payer account $s_p\in\mathcal{U}$ and specifies mode $q_p\in\{\mathsf{int},\mathsf{ext}\}$, destination $d_p$, and amount $x_p>0$.

BankCore first checks the Mandate \& Limits rules. An owner $o$ registers a mandate $M=(\sigma,o,u,a,\Pi,n)$ to let agent key $a$ spend from account $u$. Multiple agents can hold separate mandates for the same account. The policy $\Pi$ specifies a per-payment limit $\ell_{\mathrm{tx}}$, a cumulative allowance $\ell_{\mathrm{win}}$ per window of length $\omega$, a validity interval, and admissible settlement-mode and destination pairs $(q,d)$ (an explicit list or a Merkle root). Only $o$ may register or revoke $M$. In the prototype, the account address is also its owner. The mandate nonce $n$ distinguishes authorizations with otherwise identical terms.

The identifier $\mathrm{id}(M)$ is the domain-separated EIP-712 digest of the registered mandate terms: account, agent, policy, and mandate nonce. Its domain binds the chain identifier and BankCore address, which fixes the settlement asset. Registration stores the mandate under this identifier. Revocation is final, so the same identifier cannot be registered again. For payment $p$, the agent signs $I_p=(\mathrm{id}(M),q_p,d_p,x_p,n_p,\tau_p)$ using the same EIP-712 domain. Here $n_p$ is a payment nonce checked for reuse within $\mathrm{id}(M)$, and $\tau_p$ is the intent's expiration time. BankCore derives the payer from the referenced mandate, so $s_p=u$. The intent cannot select another account.

BankCore accepts an intent only if the mandate is registered and unrevoked, the signature recovers to $a$, the nonce is unused, the intent and mandate are within their validity periods, $(q_p,d_p)$ is admissible, $x_p\leq\ell_{\mathrm{tx}}$, the window allowance is not exceeded, and $b_{u,t}\geq x_p$. Acceptance consumes the payment nonce within this mandate and the allowance, which a later cancellation does not restore, before debiting the payer. All state updates precede token transfers, and the entry point is reentrancy-protected. Because intents do not bind the transaction sender, relayers can submit them, and several intents can be batched in one transaction, each validated against the state left by its predecessors (\autoref{alg:pay}).

After these checks, BankCore settles the accepted intent through the Payments block. An internal payment requires $d_p$ to be an AgentBank account for the same asset. BankCore debits $s_p$ and credits $d_p$, leaving $L_t$ and all assets unchanged. An external payment debits $s_p$ and is settled immediately if
\begin{equation}
R_t\geq O_t+x_p ,
\label{eq:immediate}
\end{equation}
This protects liquidity for earlier obligations. Withdrawals use the external path. Destination membership changes do not alter the signed mode.

\subsubsection{\flowstep{3}Queue Processing.}
If the reserve condition fails, BankCore appends the payment to the FIFO queue in the Uncovered Payment block and increases $O_t$ by $x_p$. Any address may trigger queue processing, which handles a bounded number of entries in FIFO order: an obligation $k$ whose expiration time $\tau_k$ (inherited from its payment intent) has passed is cancelled and its value returned to the payer. Otherwise, it is settled if the reserve covers its payout, and processing stops at the first unexpired obligation that cannot be settled.

\subsubsection{\flowstep{4}Impairment Handling.}
The Redemption block connects backing-asset valuations to BankCore's external payouts. BankCore reads its reserve and liabilities and obtains haircut-adjusted strategy values from TreasuryManager. The default rule processes queued obligations in FIFO order and adjusts every external payout to reported solvency. For $L_t>0$, the recovery ratio is $\kappa_t=\min\{1,\mathrm{CR}_t\}$: settlement is at par when coverage is at least one and enters impairment mode when $\mathrm{CR}_t<1$. For both immediate payments and queued settlements, BankCore transfers $\kappa_t x$ for a nominal claim $x$ and extinguishes $x$ of liabilities. Internal payments move nominal balances without transferring assets. The default has no notice period, epoch batching, or liquidation fee. With exact valuations, no haircuts, and costless recalls, it carries the pro-rata recovery of MakerDAO's emergency shutdown into continuous settlement~\cite{makerdao2026shutdown}. Adapters must report conservative, manipulation-resistant values. Delayed loss recognition can let early redeemers receive par.

The prototype implements this default rule. We evaluate three optional extensions separately in simulation. \emph{Forward pricing} delays all external payouts, including withdrawals, by a pre-committed notice period and uses the recovery ratio at settlement, following the asynchronous request model of ERC-7540. \emph{Epoch settlement} replaces FIFO priority within each epoch with pro-rata allocation of the available reserve, removing the value of ordering within that epoch~\cite{budish2015hft}. A \emph{liquidity fee} charges redeemers the discount incurred when their payouts require liquidating strategy positions, as in swing pricing~\cite{jin2022swing}. These extensions address valuation lag, transaction priority, and liquidation costs, respectively. When enabled, their rules are fixed in advance to avoid the pre-emptive runs associated with discretionary gates~\cite{cipriani2014gates,li2021liquidity}. U.S.\ money market fund reforms removed redemption gates in 2023~\cite{sec2023mmf}.

\section{Analysis}
\label{sec:analysis}

We prove that AgentBank preserves balance-sheet consistency and bounds losses from a compromised agent key by the owner-issued mandate. We derive pooled reserve requirements under correlated demand and establish how sequential par redemption concentrates losses on later redeemers. We also show when solvency-contingent redemption removes this disadvantage and why delayed loss recognition limits its protection. Full proofs are given in \autoref{app:proofs}.

\subsubsection{Account Safety.}
Internal payments should only move claims between accounts, and delegation should expose only the funds permitted by the owner's mandate. The following results formalize these two requirements.

\begin{lemma}[Conservation]
\label{lem:conservation}
Every operation in \autoref{tab:balance_sheet} satisfies $\Delta A=\Delta L+\Delta E$. Equity changes only through capital contributions, treasury gains or losses, or liability write-downs. Internal payments preserve $D$, $O$, $R$, and $T$.
\end{lemma}

\begin{proposition}[Bounded delegated loss]
\label{prop:delegation}
Let mandate $M$ on account $u$ have window allowance $\ell_{\mathrm{win}}$ and window length $\omega$, and let its session key be controlled by an adversary from time $t_c$ until the owner's revocation executes at $t_r$. Then the value debited under $M$ during $[t_c,t_r]$ is at most $\min\{\bar b_u,\ \ell_{\mathrm{win}}(\lceil (t_r-t_c)/\omega\rceil+1)\}$, where $\bar b_u$ is the total value available in $u$ during the interval, every debited payment has a destination admitted by $\Pi_M$, and no other account is debited.
\end{proposition}

\noindent This bound lets owners limit exposure to a compromised agent. Smart accounts with spending policies enforce the same bound on their funds. AgentBank does so while agents of many owners share a reserve.

\subsubsection{Reserve Requirements.}
Pooling helps because agents do not all need funds at the same time. Correlated demand limits this benefit: a shared surge still requires a shared reserve. We quantify both effects below.

\begin{proposition}[Pooled reserve]
\label{prop:pooling}
Let $X_1,\dots,X_N$ be $N$ agents' external payment demands over a replenishment horizon, and $q_\alpha$ the $\alpha$-quantile.
(i)~Per-agent prefunding at service level $\alpha$ requires $W_N=\sum_i q_\alpha(X_i)$, whereas a pooled reserve $R_N=q_\alpha(\sum_i X_i)$ covers every agent's demand with probability at least $\alpha$.
(ii)~If the $X_i$ are exchangeable with mean $m$, standard deviation $s$, and pairwise correlation $\varrho\geq 0$, and $\sum_iX_i$ is approximately Gaussian, then
$R_N/W_N\approx\bigl(m+z_\alpha s\sqrt{\varrho+(1-\varrho)/N}\bigr)/q_\alpha(X_1)$.
(iii)~If the $X_i$ are conditionally i.i.d.\ given a common factor $Z$, with $\mu(z)=\mathbb{E}[X_i\mid Z=z]$ continuous and strictly increasing and the CDF of $Z$ continuous and strictly increasing near $q_\alpha(Z)$, then $R_N/N\to\mu(q_\alpha(Z))$ as $N\to\infty$, and hence $R_N/W_N\to\mu(q_\alpha(Z))/q_\alpha(X_1)$.
\end{proposition}

\noindent Part~(ii) is the square-root law of inventory pooling~\cite{eppen1979centralization}. Part~(iii) is the large-portfolio limit behind the Basel internal-ratings formula~\cite{vasicek2002distribution,gordy2003riskfactor}: idiosyncratic demand diversifies away, whereas the common component must be reserved at its own $\alpha$-quantile. A multiplicative common factor makes aggregate demand right-skewed, so the Gaussian approximation~(ii) understates the requirement. Without a common factor, the ratio tends to $m/q_\alpha(X_1)$, which is small for bursty demand. For finite~$N$, a second-order expansion of the quantile adds the granularity adjustment of credit-portfolio models~\cite{gordy2003riskfactor,gordy2013granularity}: with $\mu_N(z)$ and $v_N(z)$ the mean and variance of $\sum_iX_i$ given $Z=z$, and $f_Z$ the density of~$Z$,
\begin{equation}
\begin{aligned}
R_N &\approx \mu_N(\zeta_\alpha)
-\frac{1}{2f_Z(\zeta_\alpha)}\,\frac{d}{dz}
\Bigl[\frac{f_Z(z)\,v_N(z)}{\mu_N'(z)}\Bigr]_{z=\zeta_\alpha}, \\
\zeta_\alpha &= q_\alpha(Z).
\end{aligned}
\label{eq:ga}
\end{equation}
whose second term does not grow with~$N$ and requires finite conditional variance.

\noindent Internalization saves reserves only while recipients retain funds. For serially uncorrelated receipts and a withdrawal fraction $\theta\in(0,1]$ per period, withdrawal variance is $\theta/(2-\theta)$ times receipt variance.

\subsubsection{Redemption Incentives.}
After an asset loss, paying early redeemers in full leaves less for those who wait. Pro-rata recovery removes this advantage once the loss is recognized, but delayed valuations leave a window for early redemption.

\begin{proposition}[Loss concentration under sequential par redemption]
\label{prop:runs}
Let $A$ and $L$ denote initial true assets and liabilities, with $A<L$. Assume their totals change only through payouts and strategy recalls are costless.
(i)~If obligations of total value $\pi L\leq A$ are settled at par before resolution, they recover in full, whereas the remaining claims recover
$\gamma(\pi)=1-\frac{1-A/L}{1-\pi}$, which is strictly decreasing in $\pi$.
(ii)~Under the default impairment rule of \autoref{sec:impairment} with $h_j=1$ and exact valuations, every claim recovers $A/L$ irrespective of the order and timing of redemption.
(iii)~If the bank meets $\mathrm{CR}\geq1$ when a loss occurs that reported values reflect only from a later time~$\tau$, every settlement before~$\tau$ pays par, whatever the haircuts, so (i) applies to the claims settled before~$\tau$. After recognition, payout ratios are bounded by the current true ratio $A_t/L_t$.
\end{proposition}

\noindent The premium for redeeming early, $(1-A/L)/(1-\pi)$, grows with the number of redeemers, which is the strategic complementarity behind bank runs~\cite{diamond1983bankRuns,goldstein2005demanddeposit}. On a public ledger, an impairment is common knowledge as soon as it is recorded, which can restore multiple equilibria~\cite{hellwig2002public}, and the position of a redemption in the queue is determined by transaction ordering, which block builders sell~\cite{daian2020flashboys}. \autoref{app:results} quantifies the resulting priority-fee race.

By \autoref{prop:runs}(iii), haircuts cannot close the window before recognition: under the coverage floor, any loss that impairs the bank exceeds the haircut buffer, which equity backs. Lending-pool share prices, for example, reflect bad debt only once it is socialized. For requests submitted after the loss, forward pricing closes this window when the notice period exceeds the recognition lag, as next-NAV pricing removes stale-price arbitrage in mutual funds~\cite{chalmers2001wildcard,zitzewitz2003arbitrage}. Internal payments can remain immediate. All external payouts share the notice period, so payment intents cannot bypass it.

\section{Evaluation}
\label{sec:evaluation}

We address four research questions:
\begin{packeditemize}
\item[$\bullet$] \textbf{RQ1:} What execution overhead does policy-constrained delegation add?
\item[$\bullet$] \textbf{RQ2:} How much idle liquidity does AgentBank save, and which mechanisms contribute?
\item[$\bullet$] \textbf{RQ3:} How do reserve policies balance treasury returns and payment availability under lending-pool stress and Treasury-bill settlement schedules?
\item[$\bullet$] \textbf{RQ4:} How do redemption rules and depositor speed affect impairment losses?
\end{packeditemize}

\subsection{Methodology}
\label{sec:method}

\subsubsection{Implementation.} We implemented BankCore, TreasuryManager, and an ERC-4626 strategy adapter in Solidity~0.8.34 (466 non-comment lines on OpenZeppelin Contracts~5.4), following the default rule in \autoref{alg:pay}. The optional redemption extensions are evaluated in simulation. The Foundry suite has 38 unit and fuzz tests, including one for \autoref{prop:delegation}, and four invariants checked over 32{,}768 random calls (balances sum to $D$, queued entries to $O$, $R$ equals the asset held, and equity is conserved). Gas is read from receipts on Anvil under the Osaka rules. All rows use one six-decimal token that implements EIP-3009 as USDC does, without its proxy and pause checks. The ERC-4337 baseline uses EntryPoint~v0.7 and a SimpleAccount whose session key is bound by the same policy as a mandate.

\subsubsection{Workload.} We simulate 1{,}000 agents, ten per owner, unless varied explicitly. Agent $i$ is active on day $t$ if $\sqrt{\varrho}Z_t+\sqrt{1-\varrho}\,\varepsilon_{i,t}$ exceeds a threshold yielding activity probability $p_i$, drawn from a beta distribution with mean 0.3 and concentration 5. The common factor $Z_t$ has AR(1) persistence 0.5, and $\varrho$ approximates pairwise demand correlation. An active agent issues a Poisson number of payments with mean $\lambda_i e^{aZ_t-a^2/2}$, where $\lambda_i\sim40\cdot\mathrm{LogNormal}(-\tfrac12,1)$. Payment sizes are log-normal with median \$0.10 and log-scale standard deviation 1. Each payment is internal with probability $\rho_{\mathrm{in}}=0.5$. Recipients withdraw a fraction $\theta=0.2$ of their balance daily, and owners replenish accounts with a one-day lag. Initial deposits $D_0$ cover owners' 99\% daily payment needs plus recipient balances.

\subsubsection{Lending-Pool Trace.} The treasury invests in the Aave~v3 USDC market on Ethereum. We replay its daily available liquidity and supply rate from 6~February~2023 to 27~October~2026 (1{,}360 days)~\cite{defillama2026yields}, scaling the bank so that its deposits equal 10\% of the market's median available liquidity, and we let the bank withdraw at most the market's available liquidity at each hour. The trace contains the freeze that followed the KelpDAO exploit: available liquidity fell from \$534~million on 18~April~2026 to below \$25{,}000 on 19--22~April, while the supply rate rose to the 12.6\% cap of the interest-rate model.

\subsubsection{Baselines and Metrics.} We compare against (i)~\emph{plain wallets}, which hold the settlement asset and earn nothing, and (ii)~\emph{yield-token wallets}, in which each agent holds the market's interest-bearing token and redeems it whenever it pays, subject to available liquidity. We report idle liquidity at a 99\% daily service level, return on assets (treasury income over deposits plus equity) under normal operation, and, during stress, the share of payment value settled immediately and the share that fails because it remains unsettled beyond its 24-hour expiration. Hourly payment simulations use FIFO settlement without asset losses, so payouts remain at par. Run experiments compare solvency-contingent redemption with the alternatives. We set $r_{\min}=2\%$, $h_j=0.98$, $\mathrm{CR}_{\min}=1$, and initial equity $E_0=0.02D_0$, rebalancing every six hours with a 10\% tolerance. Results are averaged over five workload seeds.

\subsection{RQ1: Delegation Overhead}
\label{sec:rq1}

\autoref{tab:gas} compares AgentBank's steady-state gas with ERC-20, EIP-3009 (x402), and policy-equivalent ERC-4337 transfers. An internal payment costs 61.4k gas, including 26.5k for delegation over an owner's ledger transfer (34.8k). This covers EIP-712 hashing, signature recovery, four mandate reads, window and nonce updates, and calldata. A proof over 1{,}024 destinations adds 8.3k. Internal payments cost less than EIP-3009, which stores each random nonce in a new slot, and 56\% of a policy-equivalent UserOperation. Batching $k=50$ lowers the cost to 40.0k (1.17 times an ERC-20 transfer), compared with 46.6k for EIP-3009 and 72.9k for UserOperations.

External payments add 23.6k for asset transfer and 23.1k for strategy valuations required by $\kappa_t$. Batching reduces this overhead to 11.1k. Cached valuations cost one storage read but allow par redemptions until refreshed (\autoref{prop:runs}(iii)). Queued payments cost 86--115k when reserves fall short.

\begin{table}[t]
\centering
\caption{Gas per payment and calldata bytes per single-payment transaction.}
\label{tab:gas}
\scriptsize
\begin{tabular*}{\textwidth}{@{\extracolsep{\fill}}lcccc@{}}
\toprule
 & \multicolumn{3}{c}{\textbf{Gas per payment}} & \\
\cmidrule(lr){2-4}
\textbf{Operation} & \textbf{single} & $k=10$ & $k=50$ & \textbf{Bytes} \\
\midrule
\textbf{ERC-20} transfer from an agent wallet & 34{,}313 & -- & -- & 68 \\
\textbf{EIP-3009} \texttt{transferWithAuthorization} (x402) & 65{,}222 & 48{,}524 & 46{,}587 & 292 \\
\textbf{ERC-4337} session-key transfer & 109{,}771 & 75{,}840 & 72{,}929 & 900 \\
\midrule
\textbf{AgentBank internal payment} & 61{,}353 & 41{,}814 & 40{,}031 & 324 \\
\quad with destination proof (1{,}024 destinations) & 69{,}646 & -- & -- & 644 \\
\textbf{AgentBank external payment}, immediate & 108{,}095 & 55{,}797 & 51{,}111 & 324 \\
\quad without strategy valuation ($J=0$) & 84{,}955 & -- & -- & 324 \\
\textbf{AgentBank external payment}, queued & 114{,}509 & 86{,}323 & -- & 324 \\
\textbf{Queue settlement}, per entry & 77{,}921 & 22{,}719 & -- & 36 \\
\bottomrule
\end{tabular*}
\end{table}

\subsection{RQ2: Liquidity Efficiency}
\label{sec:rq2}

\begin{figure}[t]
\centering
\includegraphics[width=\textwidth]{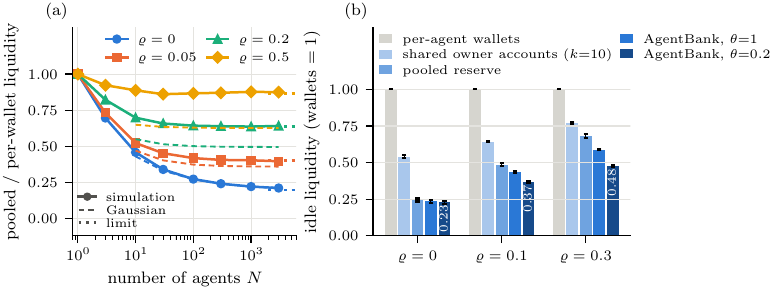}
\caption{Liquidity savings: \textbf{(a)}~reserve scaling; \textbf{(b)}~pooling and internalization.}
\label{fig:liquidity}
\end{figure}

\autoref{fig:liquidity}(a) varies the population from 1 to 3{,}000 agents and correlations $\varrho=0$, 0.05, 0.1, 0.2, 0.3, and 0.5 to measure pooled reserve requirements. Without a common factor, the requirement falls to 22\% of per-agent prefunding at $N=1{,}000$ and approaches the limit $m/q_\alpha(X_1)=0.20$, as predicted. Under correlated demand, the requirement plateaus far earlier and at a higher level: 0.40, 0.64, and 0.88 of per-agent prefunding for $\varrho=0.05$, $0.2$, and $0.5$. The Gaussian approximation understates these plateaus by up to 29\%, whereas the large-pool limit of \autoref{prop:pooling}(iii) matches the simulation to within 1.5\%. For finite pools, \autoref{eq:ga} reduces error below 2\% from 30 homogeneous or 100 heterogeneous agents. The limit alone errs by up to 12\% and 8\% (\autoref{tab:ga}).

\autoref{fig:liquidity}(b) decomposes the saving for heterogeneous agents. Letting ten agents of one owner share an account already reduces idle liquidity by 23--46\%. Pooling across all owners reduces it by 32--76\%, and internalization with recipient retention ($\theta=0.2$) adds a further 2--21 percentage points, so that AgentBank requires 0.23, 0.37, and 0.48 of per-agent prefunding for $\varrho=0$, $0.1$, and $0.3$, a saving of 52--77\%. Relative to the strongest wallet-based alternative, one smart account shared by an owner's agents, the saving is 38--58\%. As in tiered payment systems~\cite{lasaosa2008tiering}, pooling contributes most of the saving. Internalization matters most under high correlation, as retained balances smooth common shocks. With immediate withdrawals ($\theta=1$), it saves only 1--9 percentage points.

With $\varrho=0.1$, raising $\rho_{\mathrm{in}}$ from 0 to 0.9 lowers idle liquidity from 0.48 to 0.31 of per-agent prefunding and reduces daily token transfers from about 12{,}200 to 1{,}320, because internal payments transfer none and the 100 recipients withdraw aggregated balances. Recipient concentration makes high $\rho_{\mathrm{in}}$ attainable: in x402 on Base, the largest recipient receives 23\% of settlements and the three largest more than half~\cite{ling2026agenticCommerceMeasurement}. Under a Zipf law with exponent~1 over $10^4$ recipients, onboarding the ten most popular internalizes 30\% of volume. Recipients rarely pay onward (reciprocity coefficient 0.025~\cite{ling2026agenticCommerceMeasurement}), so their withdrawal rate $\theta$ determines the internalization benefit. Heavy-tailed demand raises the saving to 79--89\%.

\subsection{RQ3: Efficiency--Resilience Trade-off}
\label{sec:rq3}

\begin{figure}[t]
\centering
\includegraphics{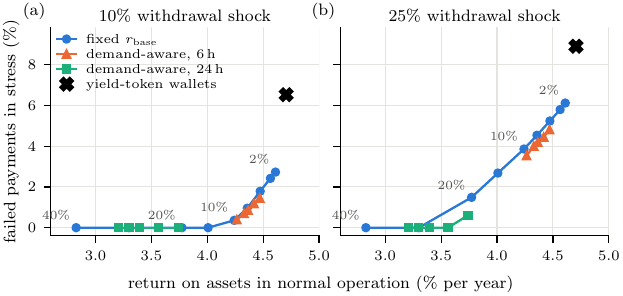}
\caption{Efficiency--resilience frontier under \textbf{(a)}~10\% and \textbf{(b)}~25\% withdrawal shocks.}
\label{fig:frontier}
\end{figure}

We compare fixed reserve ratios of 2--40\% with demand-aware policies using quantiles of 0.5--0.99, horizons of $\Delta\in\{6,24\}$ hours, and $W=14$ days of history. Under normal operation, a small reserve suffices because the treasury can recall liquidity every six hours: with $r_{\mathrm{base}}=5\%$, 99.2\% of external payment value settles immediately and the return on assets is 4.47\% per year, compared with a mean supply rate of 4.71\%. Each further 10~percentage points of reserve cost about 0.47~points of return.

The freeze changes the picture. We evaluate 18~April--2~May~2026, with withdrawals of 10\% or 25\% of $D_0$ arriving over 24~hours from noon on 18~April and owners halving their top-ups for 5.5~days, reflecting a flight to safety (\autoref{fig:frontier}). Agents that hold yield-bearing tokens directly earn the full supply rate, but 6.5\% (10\% shock) and 8.9\% (25\% shock) of their payment value fails, and even without a shock 4.8\% fails while redemptions are blocked. AgentBank gives up part of this return in exchange for far fewer failures: with $r_{\mathrm{base}}=10\%$, failures fall to 0.4\% and 3.9\% at a return of 4.24\%. With $r_{\mathrm{base}}=15\%$, no payment fails under the 10\% shock (return 4.01\%), and $r_{\mathrm{base}}=30\%$ eliminates failures under the 25\% shock (return 3.30\%). The demand-aware policy lies on the same frontier. With a 24-hour horizon it holds 21--34\% and avoids nearly all failures, although it sees only AgentBank's flows and cannot predict external freezes.

In the freeze, failures of yield-token wallets reach 18\% of deposits within a day, whereas AgentBank keeps settling from its reserve and owners' deposits (\autoref{fig:stress_series}): the reserve converts a venue freeze into a bounded queueing delay at a cost that the operator sets.

These conclusions survive partial capture of the pool's liquidity, a doubling of external turnover, diversification across venues, and a replay of the Solend freeze of June~2022 (\autoref{app:results}). Freezes recur: 64 stablecoin lending pools saw 46 since 2022, and the April~2026 freeze reached 14 pools on seven chains.

\subsubsection{Compliant Treasury.} The GENIUS Act admits tokenized Treasury bills as stablecoin reserves but forbids rehypothecation~\cite{genius2025act}, which excludes lending pools. A Treasury-bill fund never freezes but redeems only on business days, so the weekend becomes the binding stress (\autoref{fig:calendar}): settling 99\% of payment value immediately requires a fixed reserve of 20\% instead of 5\%, and keeping failures below 0.05\% under a 10\% shock at the worst time of the week requires 30\% (same-day redemption) or 40\% (next-day), which retains 72\% or 62\% of the strategy yield. Compliance thus exchanges venue risk for calendar risk.

\subsection{RQ4: Redemption Rules and Depositor Speed}
\label{sec:rq4}

\begin{figure}[t]
\centering
\includegraphics{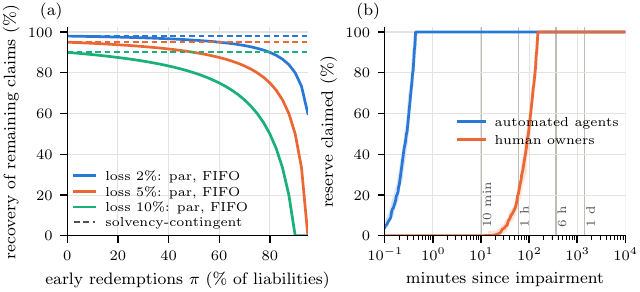}
\caption{Redemption incentives: \textbf{(a)}~remaining-claim recovery; \textbf{(b)}~reserve capture.}
\label{fig:runs}
\end{figure}

We now switch from liquidity-freeze replays to depositor models of asset impairment. \autoref{fig:runs}(a) illustrates \autoref{prop:runs}(i)--(ii). With a 5\% asset loss, depositors who wait recover 90\% if half of all liabilities have been redeemed at par before them, and 50\% if 90\% have. With exact values, no haircuts, and no liquidation costs, solvency-contingent redemption gives everyone 95\%.

\autoref{fig:runs}(b) shows why the redemption rule must be mechanical. With log-normal reaction delays (median one minute for automated agents, six hours for human owners), automated depositors claim the entire reserve within 27~seconds of an impairment becoming public. Human owners claim 21\% of it within an hour and all of it within 2.6~hours. A discretionary response, such as pausing withdrawals, therefore arrives after an automated run has completed, and a gate that is known in advance invites pre-emptive redemptions~\cite{cipriani2014gates,li2021liquidity}.

In equilibrium, FIFO par settlement leaves the loss to human owners or, once automated claims exceed the assets, starts a priority-fee race that costs more than the loss. With prompt loss recognition and costless liquidation, solvency-contingent redemption removes the run, and epoch-based pro-rata settlement removes the value of priority within an epoch.

If a 5\% loss reaches reported values one minute late, however, human owners recover 93.4\% instead of 95\%, and after 30~minutes 90.2\%, as under FIFO, whatever the haircut. With private, noisy loss signals, the FIFO equilibrium is unique, and runs start about three signal standard deviations before a run at par would cause insolvency. Costly liquidation also permits runs under solvency-contingent redemption. The liquidity fee removes this modeled equilibrium.

\section{Discussion}
\label{sec:discussion}

\subsubsection{Economic Design.} Owner pooling lowers what owners must deposit, whereas bank pooling and internalization lower what the bank holds idle. The value of the latter accrues to whoever receives the treasury return, which the evaluated operator keeps as compensation for first-loss capital. Passing part of it to depositors would raise adoption but tighten the frontier of \autoref{fig:frontier}, and a higher return on equity strengthens the incentive to reduce reserves~\cite{goel2026stablecoinRegulation}.

\subsubsection{Regulation.} AgentBank accepts deposits and transforms liquidity, which most jurisdictions regulate: the U.S.\ GENIUS Act requires cash-like reserves, forbids their rehypothecation, and prohibits interest to holders~\cite{genius2025act}, and MiCA prohibits interest on e-money tokens~\cite{mica2023regulation}. We do not resolve whether AgentBank balances are payment stablecoins, e-money, or deposits. Treasury-bill funds without yield to holders remain viable but need larger reserves unless 24/7 conversion facilities~\cite{circle2024buidl} bridge business days. Alternatively, a chartered bank could operate AgentBank as a tokenized-deposit system~\cite{jpmorgan2025jpmd} with verifiable liabilities~\cite{dagher2015provisions,ji2021gpol}.

\subsubsection{Limitations.} The prototype is unaudited and measured with a simplified token. The workload is synthetic, because measured x402 activity is dominated by manufactured settlements~\cite{ling2026agenticCommerceMeasurement}. These data inform only recipient concentration, and our robustness analysis covers heavier tails~\cite{bimpikis2016heavytailed} but not demand from deployed agents. Liquidity results are ratios and thus independent of payment sizes. The treasury replays use daily liquidity. Strategy losses enter only the run model, depegs are outside the model, the run models assume risk-neutral depositors, and insolvency resolution beyond the impairment mode remains future work.

\section{Conclusion}
\label{sec:conclusion}
We propose AgentBank, a smart-contract banking architecture for AI-agent payments that separates delegated spending from the management of pooled backing assets. The design combines owner-issued mandates, internal settlement, and reserve-constrained treasury allocation with explicit redemption rules for asset impairment. We establish its spending bounds and liquidity requirements, and use a Solidity prototype and trace-driven evaluation to characterize the trade-off between capital efficiency and payment resilience.

\clearpage
\bibliographystyle{splncs04}
\bibliography{ref}

\appendix

\section{Notation}
\label{app:notation}

\begin{table}[H]
\centering
\caption{Notation.}
\label{tab:notation}
\scriptsize
\renewcommand{\arraystretch}{1.15}
\setlength{\tabcolsep}{3pt}
\begin{tabular*}{\textwidth}{@{\extracolsep{\fill}}clcl@{}}
\toprule
\textbf{Symbol} & \multicolumn{1}{c}{\textbf{Meaning}} & \textbf{Symbol} & \multicolumn{1}{c}{\textbf{Meaning}} \\
\midrule
$\sigma$ & Settlement asset & $A^{\mathrm{eff}}_t$ & Effective assets \\
$b_{u,t}$ & Account balance & $\mathrm{CR}_t$ & Coverage ratio \\
$D_t,\ O_t$ & Deposits, queued obligations & $r_{\min},\ \mathrm{CR}_{\min}$ & Reserve floor, coverage floor \\
$L_t$ & Total liabilities & $r_{\mathrm{base}}$ & Baseline reserve ratio \\
$R_t$ & Settlement reserve & $\alpha$ & Service-level quantile \\
$T_{j,t}$ & Strategy position & $W$ & Look-back duration (days) \\
$A_t,\ E_t$ & Assets, equity & $\Delta$ & Planning horizon (hours) \\
$M,\ \Pi$ & Mandate, policy & $\mathrm{id}(M)$ & Mandate identifier \\
$I_p,\ \tau_p$ & Intent, expiration time & $\rho_{\mathrm{in}}$ & Internal payment ratio \\
$\ell_{\mathrm{tx}}$ & Per-payment limit & $\varrho$ & Demand correlation \\
$\ell_{\mathrm{win}},\ \omega$ & Window allowance, duration & $\theta$ & Recipient withdrawal fraction \\
$h_j$ & Valuation factor & $\kappa_t$ & Impairment recovery ratio \\
\bottomrule
\end{tabular*}
\end{table}

\Needspace{31\baselineskip}
\section{Algorithms}
\label{app:algorithms}

\autoref{alg:pay} summarizes payment execution. Beyond \autoref{tab:gas}, an owner ledger transfer costs 34{,}809 gas. Deposits and withdrawals cost 58{,}430 and 82{,}294 gas, respectively. Mandate registration and revocation cost 118{,}209 and 28{,}256 gas. Treasury deployment and recall cost 112{,}894 and 86{,}079 gas.

\begin{algorithm}[H]
\caption{BankCore payment execution and queue processing}
\label{alg:pay}
\footnotesize
\algrenewcommand\algorithmicindent{1em}
\noindent\textbf{Input:} Asset $\sigma$, intent $I_p=(\mathrm{id}(M),q,d,x,n_p,\tau)$ and signature.
\par\noindent Each intent or queue-processing call sets $\kappa=\min\{1,A^{\mathrm{eff}}/L\}$ for $L>0$ before changing liabilities. Transfers reduce $R$. State indices are omitted.
\par\smallskip
\noindent\textbf{Validate intent}
\begin{algorithmic}[1]
  \State $M=(\sigma,o,u,a,\Pi,n)\gets$ owner-registered mandate at $\mathrm{id}(M)$
  \State $s_p\gets u$; $t\gets$ block timestamp
  \State \Req $M$ not revoked; signature recovers to $a$; $n_p$ unused under $\mathrm{id}(M)$
  \State \Req $t\leq\tau$ and $t$ within the validity period of $\Pi$
  \State \Req $(q,d)$ admissible under $\Pi$; if $q=\mathsf{int}$, $d$ is an account for $\sigma$
  \State Reset window usage if the window has advanced
  \State \Req $x\leq\ell_{\mathrm{tx}}$, $\mathrm{used}+x\leq\ell_{\mathrm{win}}$, and $b_u\geq x$
  \State Mark $n_p$ used under $\mathrm{id}(M)$; $\mathrm{used}\gets\mathrm{used}+x$; $b_u\gets b_u-x$ \label{alg:debit}
  \algstore{validatedpayment}
\end{algorithmic}
\smallskip
\noindent\begin{minipage}[t]{0.485\linewidth}
\textbf{Settle accepted intent}
\begin{algorithmic}[1]
  \algrestore{validatedpayment}
  \If{$q=\mathsf{int}$}
    \State $b_d\gets b_d+x$
  \Else
    \State $D\gets D-x$
    \If{$R\geq O+x$}
      \State Transfer $\kappa x$ of $\sigma$ to $d$
    \Else
      \State $O\gets O+x$
      \State $e\gets(u,d,x,\tau)$
      \State Append $e$ to the queue
    \EndIf
  \EndIf
\end{algorithmic}
\end{minipage}\hfill
\begin{minipage}[t]{0.485\linewidth}
\textbf{ProcessQueue}$(k)$ \textit{(any caller)}
\begin{algorithmic}[1]
  \For{up to $k$ queued entries}
    \State $(u,d,x,\tau)\gets$ queue head
    \If{$t>\tau$}
      \State Pop head; $O\gets O-x$
      \State $b_u\gets b_u+x$; $D\gets D+x$
    \ElsIf{$R\geq\kappa x$}
      \State Pop head; $O\gets O-x$
      \State Transfer $\kappa x$ of $\sigma$ to $d$
    \Else
      \State \textbf{break}
    \EndIf
  \EndFor
\end{algorithmic}
\end{minipage}
\par\vspace{4pt}
\end{algorithm}

\section{Proofs}
\label{app:proofs}
\suppressfloats[t]

\begin{proof}[\autoref{lem:conservation}]
By inspection of \autoref{tab:balance_sheet}: in each row, $\Delta A=\Delta R+\Delta T$ and $\Delta L=\Delta D+\Delta O$, and the difference equals the entry in the $\Delta E$ column. Settling a nominal claim $x$ at recovery ratio $\kappa$ gives $\Delta A=-\kappa x$, $\Delta L=-x$, and $\Delta E=(1-\kappa)x$. This write-down vanishes at par. The remaining equity changes are capital contributions and treasury gains or losses. An internal payment changes two balances by $\mp x$ and therefore leaves $D$ and all other quantities unchanged. \qed
\end{proof}

\begin{table}[t]
\centering
\begin{threeparttable}
\caption{Balance-sheet effects of AgentBank operations.}
\label{tab:balance_sheet}
\scriptsize
\setlength{\tabcolsep}{4pt}
\begin{tabular}{@{}lccccccc@{}}
\toprule
\textbf{Operation} & $\Delta D$ & $\Delta O$ & $\Delta L$ & $\Delta R$ & $\Delta T$ & $\Delta A$ & $\Delta E$ \\
\midrule
\textbf{Deposit}                     & $+x$ & $0$  & $+x$ & $+x$ & $0$  & $+x$ & $0$ \\
\textbf{Capital contribution}        & $0$  & $0$  & $0$  & $+x$ & $0$  & $+x$ & $+x$ \\
\textbf{Internal payment}            & $0$  & $0$  & $0$  & $0$  & $0$  & $0$  & $0$ \\
\textbf{External payment, immediate} & $-x$ & $0$  & $-x$ & $-\kappa x$ & $0$  & $-\kappa x$ & $(1-\kappa)x$ \\
\textbf{External payment, queued}    & $-x$ & $+x$ & $0$  & $0$  & $0$  & $0$  & $0$ \\
\textbf{Queue settlement}            & $0$  & $-x$ & $-x$ & $-\kappa x$ & $0$  & $-\kappa x$ & $(1-\kappa)x$ \\
\textbf{Queue cancellation}          & $+x$ & $-x$ & $0$  & $0$  & $0$  & $0$  & $0$ \\
\textbf{Treasury deploy / recall}    & $0$  & $0$  & $0$  & $\mp x$ & $\pm x$ & $0$  & $0$ \\
\textbf{Treasury gain / loss}        & $0$  & $0$  & $0$  & $0$  & $\pm y$ & $\pm y$ & $\pm y$ \\
\bottomrule
\end{tabular}
\begin{tablenotes}[flushleft]
\scriptsize
\item $x$: nominal operation value. $y$: treasury gain or loss. $T=\sum_jT_j$.
\item $\kappa$: recovery ratio before settlement. $\kappa=1$ at par. A discounted payout writes off $(1-\kappa)x$ of liabilities, increasing book equity.
\end{tablenotes}
\end{threeparttable}
\end{table}

\begin{proof}[\autoref{prop:delegation}]
Every accepted intent obtains its payer $s_p=u$ from the registered mandate identified by $\mathrm{id}(M)$ and debits that account (\autoref{alg:pay}, line~\ref{alg:debit}), so no other account is debited, and the admissibility check restricts destinations to $\Pi_M$. Allowance is consumed at acceptance and never restored, so the accepted value within each window is at most $\ell_{\mathrm{win}}$. An interval of length $t_r-t_c$ intersects at most $\lceil (t_r-t_c)/\omega\rceil+1$ windows. The payment nonce is consumed within $\mathrm{id}(M)$, whose registration cannot be repeated, and the EIP-712 domain binds the chain and BankCore deployment. Thus an intent cannot be accepted twice or under another deployment. Finally, debits never exceed the balance of $u$. \qed
\end{proof}

\begin{proof}[\autoref{prop:pooling}]
(i)~If $\sum_iX_i\leq R_N$, every external payment in the horizon satisfies \autoref{eq:immediate}, so each agent is fully served on this event, which has probability at least $\alpha$. (ii)~$\mathrm{Var}(\sum_iX_i)=Ns^2(1+(N-1)\varrho)$, and the Gaussian quantile is $Nm+z_\alpha\sqrt{\mathrm{Var}(\sum_iX_i)}$. Dividing by $Nq_\alpha(X_1)$ yields the expression. (iii)~Conditionally on $Z$, the strong law of large numbers gives $\frac1N\sum_iX_i\to\mu(Z)$ almost surely, hence in distribution. The assumptions on $\mu$ and the CDF of $Z$ make the limiting CDF continuous and strictly increasing near $q_\alpha(\mu(Z))=\mu(q_\alpha(Z))$. Thus the quantiles of $\frac1N\sum_iX_i$ converge to this value~\cite{gordy2003riskfactor}. \qed
\end{proof}

\subsubsection{Granularity Adjustment.} Write $\sum_iX_i=\mu_N(Z)+\varepsilon$ with $\mathbb{E}[\varepsilon\mid Z]=0$ and $\mathrm{Var}(\varepsilon\mid Z)=v_N(Z)$. A second-order expansion of the $\alpha$-quantile in the idiosyncratic term gives $q_\alpha(\sum_iX_i)\approx y_\alpha-\frac{1}{2g(y_\alpha)}\frac{d}{dy}[g(y)w(y)]_{y=y_\alpha}$, where $g$ is the density of $\mu_N(Z)$, $y_\alpha$ its $\alpha$-quantile, and $w(\mu_N(z))=v_N(z)$~\cite{gordy2003riskfactor,gordy2013granularity}. Since $\mu_N$ is increasing, $y_\alpha=\mu_N(\zeta_\alpha)$, $g(\mu_N(z))=f_Z(z)/\mu_N'(z)$, and $dy=\mu_N'(z)\,dz$, which yields \autoref{eq:ga}. In our workload, given $Z=z$, agent~$i$ is active with probability $p_i(z)=\Phi\bigl((\sqrt{\varrho}z-c_i)/\sqrt{1-\varrho}\bigr)$ for its activity threshold $c_i$, and then issues a Poisson number of payments with mean $\Lambda_i(z)B_i$, where $\Lambda_i(z)=\lambda_ie^{az-a^2/2}$, $B_i$ is a burst multiplier with mean one ($B_i=1$ without bursts), and sizes~$S$. Hence $\mu_N(z)=\sum_ip_i\Lambda_i\mathbb{E}[S]$ and $v_N(z)=\sum_i\bigl[p_i(\Lambda_i\mathbb{E}[S^2]+\Lambda_i^2\mathbb{E}[S]^2\mathbb{E}[B_i^2])-(p_i\Lambda_i\mathbb{E}[S])^2\bigr]$, which is finite only if $\mathbb{E}[B_i^2]<\infty$ (Lomax tail index greater than~2).

\begin{proof}[Retention]
If recipients withdraw a fraction $\theta$ of their balance per period, the recipient balance obeys $M_{t+1}=(1-\theta)M_t+I_t$, so the withdrawals are $W_t=\theta M_t=\theta\sum_{k\geq0}(1-\theta)^kI_{t-1-k}$. For serially uncorrelated receipts,
\[
\mathrm{Var}(W_t)=\theta^2\mathrm{Var}(I)\textstyle\sum_{k\geq0}(1-\theta)^{2k}=\frac{\theta^2}{1-(1-\theta)^2}\mathrm{Var}(I)=\frac{\theta}{2-\theta}\mathrm{Var}(I). \qquad\qed
\]
\end{proof}

\begin{proof}[\autoref{prop:runs}]
(i)~Par settlement of $\pi L$ leaves assets $A-\pi L\geq0$ for liabilities $(1-\pi)L$, which yields $\gamma(\pi)$. Its derivative with respect to $\pi$ is $-(1-A/L)/(1-\pi)^2<0$. (ii)~In impairment mode with $h_j=1$ and exact valuations, $\kappa=A/L$, and settling a claim $x$ leaves $(A-\kappa x)/(L-x)=\kappa$, so the ratio, and hence every subsequent payout ratio, is invariant. (iii)~Until~$\tau$, reported positions keep their pre-loss values, so the reported effective assets $\hat A^{\mathrm{eff}}$ satisfy $\hat A^{\mathrm{eff}}\geq L$ when the loss occurs. A par settlement of $x$ lowers $\hat A^{\mathrm{eff}}$ and $L$ by $x$, and a recall of $z$ from strategy~$j$ raises $\hat A^{\mathrm{eff}}$ by $(1-h_j)z\geq0$, so $\hat A^{\mathrm{eff}}-L\geq0$ and $\kappa=1$ until~$\tau$, irrespective of $h_j$. The claims settled before~$\tau$ are therefore paid at par, as in~(i). From~$\tau$ on, valuations are exact, so $\hat A^{\mathrm{eff}}_t=R_t+\sum_jh_jT_{j,t}\leq A_t$ and $\kappa_t\leq A_t/L_t$. Settling a claim $x<L_t$ yields $(A_t-\kappa_t x)/(L_t-x)\geq A_t/L_t$. Thus current true asset coverage does not decrease under the stated assumptions. Finally, $\mathrm{CR}\geq1$ with exact valuations implies $E\geq\sum_j(1-h_j)T_j$, so a loss within the haircut buffer leaves $A\geq L$: every loss that impairs the bank exceeds it. \qed
\end{proof}

\section{Additional Results and Robustness}
\label{app:results}

\setcounter{topnumber}{1}
\setcounter{bottomnumber}{1}
\setcounter{totalnumber}{2}

\begin{figure}[h]
\centering
\includegraphics[width=0.85\textwidth]{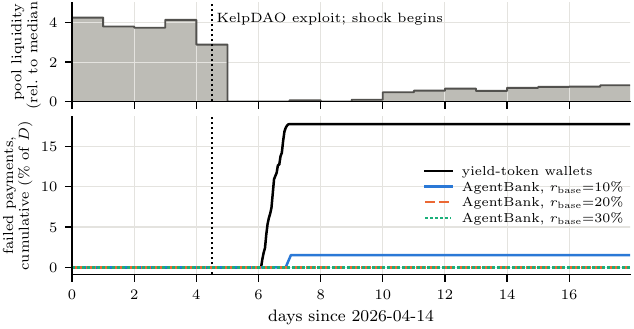}
\caption{April~2026 freeze: liquidity (top); cumulative payment failures (bottom).}
\label{fig:stress_series}
\end{figure}

We test reserve estimation, liquidity stress, and redemption incentives. Unless varied below, settings match the main evaluation. Failures refer to the April~2026 freeze with a 10\% withdrawal shock. Results average five workload seeds, or 20 populations for run experiments.

\subsubsection{Heavy-Tailed Demand.} 
\begin{figure}[tbp]
\centering
\includegraphics{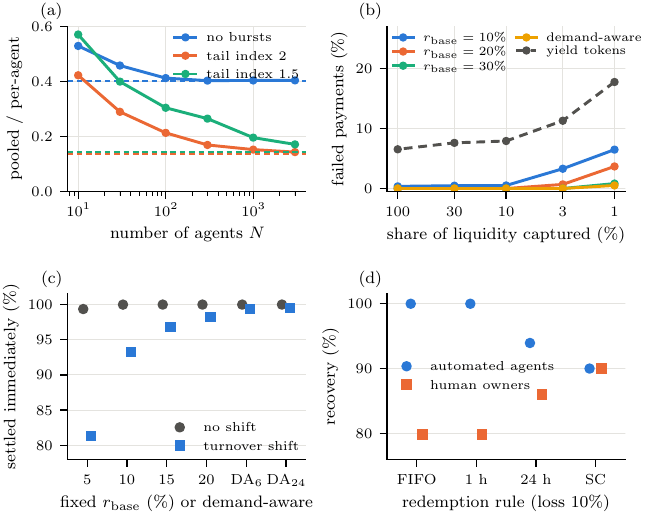}
\caption{Robustness: \textbf{(a)}~bursts; \textbf{(b)}~liquidity capture; \textbf{(c)}~turnover; \textbf{(d)}~redemption.}
\label{fig:robustness}
\end{figure}

Daily payment intensity is multiplied by independent, mean-one Lomax factors with tail indices 3, 2, or 1.5. We also replace Gaussian common-factor innovations with Student-$t$ innovations with 5 or 3 degrees of freedom. At $N=1{,}000$, the large-pool limit evaluated at the factor's own quantile is within 0.6--5.5\% of simulation, versus $-25\%$ to $+15\%$ error for the Gaussian approximation. Bursts raise per-agent prefunding from 5.2 to about 15 times mean daily demand and pooling savings from 52--77\% to 79--89\%. Convergence is slow: at tail index 1.5, the limit understates reserves by 27\% at $N=1{,}000$ and 17\% at $N=3{,}000$ (\autoref{fig:robustness}(a)).

\subsubsection{Finite-Pool Reserve Rule.} \autoref{tab:ga} compares the large-pool limit with the granularity adjustment in \autoref{eq:ga}. At $N=100$, the adjustment reduces error from 3--8\% to below 2\% for heterogeneous agents and from 21\% to 0.2\% for tail-index-3 bursts. It overshoots near $N=10$ and requires finite conditional variance. For tail indices at or below 2, the demand-aware policy must use empirical quantiles. In deployment, the conditional moments would be estimated from bank flows and a market-wide activity index.

\begin{table}[tbp]
\centering
\begin{threeparttable}
\caption{Reserve estimation error (\%) relative to the simulated 99\% quantile.}
\label{tab:ga}
\scriptsize
\setlength{\tabcolsep}{3pt}
\begin{tabular*}{\textwidth}{@{\extracolsep{\fill}}rrrrrrrrrrrrr@{}}
\toprule
 & \multicolumn{4}{c}{\textbf{Homogeneous}} & \multicolumn{4}{c}{\textbf{Heterogeneous}} & \multicolumn{4}{c}{\textbf{Bursts}} \\
\cmidrule(lr){2-5}\cmidrule(lr){6-9}\cmidrule(lr){10-13}
 & \multicolumn{2}{c}{$\varrho=0.05$} & \multicolumn{2}{c}{$\varrho=0.2$} & \multicolumn{2}{c}{$\varrho=0.05$} & \multicolumn{2}{c}{$\varrho=0.3$} & \multicolumn{2}{c}{\textbf{tail 3}} & \multicolumn{2}{c}{\textbf{tail 2.5}} \\
{\bfseries $N$} & \textbf{limit} & \textbf{GA} & \textbf{limit} & \textbf{GA} & \textbf{limit} & \textbf{GA} & \textbf{limit} & \textbf{GA} & \textbf{limit} & \textbf{GA} & \textbf{limit} & \textbf{GA} \\
\midrule
10 & $-23.4$ & 9.4 & $-8.9$ & 1.8 & -- & -- & -- & -- & $-60.5$ & 40.9 & $\boldsymbol{-64.3}$ & \textbf{78.1} \\
30 & $-11.5$ & 1.1 & $-3.3$ & 0.4 & $-18.4$ & 5.7 & $-4.1$ & 0.8 & $-41.7$ & 8.3 & $-47.3$ & 23.0 \\
100 & $-4.4$ & $-0.3$ & $-1.1$ & 0.1 & $-7.7$ & $-0.3$ & $-3.0$ & $-1.6$ & $\boldsymbol{-20.6}$ & $\boldsymbol{-0.2}$ & $-25.3$ & 4.6 \\
300 & $-1.4$ & 0.0 & $-0.5$ & $-0.1$ & $-3.5$ & $-0.5$ & $-1.1$ & $-0.6$ & $-8.7$ & $-0.8$ & $-12.2$ & $-0.5$ \\
1{,}000 & $-0.7$ & $-0.3$ & $-0.1$ & 0.0 & $-1.3$ & $-0.1$ & 0.3 & 0.5 & $-3.1$ & $-0.7$ & $-3.9$ & $-0.1$ \\
\bottomrule
\end{tabular*}
\begin{tablenotes}[flushleft]
\scriptsize
\item Limit: large-pool limit. GA: granularity-adjusted rule.
\item Bold pairs show improvement at $N=100$ (tail 3) and overshoot at $N=10$ (tail 2.5). Smaller absolute error is better.
\end{tablenotes}
\end{threeparttable}
\end{table}

\subsubsection{Partial Capture of Pool Liquidity.} The baseline allows access to all available pool liquidity. The simulated bank's pro-rata share is about 1\% of supply. Reducing capture from 100\% to 1\% raises failures from zero to 5.1\% at $r_{\mathrm{base}}=15\%$ and to 0.8\% at 30\%, while 40\% still avoids failures (\autoref{fig:robustness}(b)). Yield-token wallets fail on 17.7\%. The 24-hour demand-aware policy holds about one-third of deposits and fails on 0.5\%. Normal-operation returns are unchanged.

\subsubsection{Turnover Shift.} On 1~January~2025, we reduce the internal share from 0.5 to 0.2 and raise $\theta$ from 0.2 to 1, keeping owners' prefunding unchanged. Deposits fall to 42\% of their initial level, roughly doubling external turnover. Fixed reserves of 5\%, 10\%, and 15\% then settle 81\%, 93\%, and 97\% immediately (\autoref{fig:robustness}(c)). At $r_{\mathrm{base}}=10\%$, freeze failures rise from 0.6\% to 1.4\%. The 6-hour demand-aware policy raises its reserve from 10\% to 27\% and maintains 99.4\% immediate settlement, earning 2.7\% versus 3.4\% for the fixed 10\% policy. A 24-hour horizon holds 75\% and earns 1.0\%. Adaptive reserves track faster outflows at lower yield.

\subsubsection{Diversification across Venues.} We split assets equally between Aave~v3 on Ethereum and Base, Aave and Compound~v3 on Ethereum, or all three. Over 10~March~2024--27~October~2026, each split avoids failures under both shocks even with 5\% reserves, earning 4.27--4.36\% versus 4.30\% for Aave alone at 10\% reserves. Base froze for one day and Compound did not freeze. Yield-token wallets on these venues also avoided failures. Diversification substitutes for reserves in this episode, but correlated freezes and smaller venues' liquidity limit this protection.

\subsubsection{Run Equilibrium.} \autoref{fig:robustness}(d) uses 5{,}000 depositors with log-normal balances (log-scale standard deviation 1.5), half automated and half human. Log-normal reaction delays have log-scale standard deviation 1 and medians of one minute and six hours. After loss $\ell$, depositors redeem when expected payout minus a \$1 transaction cost exceeds recovery from waiting. Iterating best responses from ``nobody redeems'' finds the smallest equilibrium. Under FIFO par settlement with $\ell=10\%$, automated agents recover 100\%, humans 79.8\%, and 20\% of humans lose over half their balance. One-hour epochs change nothing because agents fill the first epoch. A 24-hour epoch raises human recovery to 86.0\%. Solvency-contingent redemption yields no run and 90\% recovery. Raising costs to \$20 changes recovery by at most about one percentage point. With 80\% automated depositors and $\ell=5\%$, human FIFO recovery is 75\%.

\clearpage
\begin{figure}[H]
\centering
\includegraphics{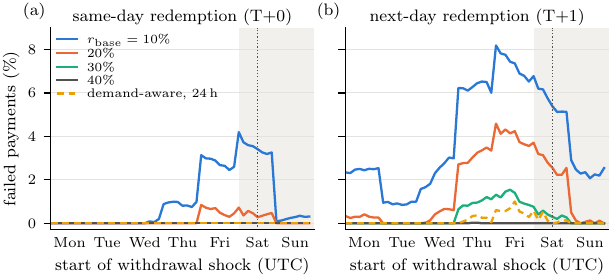}
\caption{Treasury-bill settlement risk: \textbf{(a)}~same-day and \textbf{(b)}~next-day redemption.}
\label{fig:calendar}
\end{figure}

\subsubsection{Treasury-Bill Fund.} We replace the lending pool with a tokenized Treasury-bill fund consistent with the reserve rules of the GENIUS Act~\cite{genius2025act}. The fund has no liquidity limit. T+0 requests before 18:00~UTC settle at 21:00 that business day, with later requests rolling forward. T+1 requests settle on the following business day. Weekends and U.S.\ federal holidays are excluded. Assets in transit accrue yield and count toward coverage. We report retained strategy yield, independent of its assumed constant rate. Instant redemption, as offered by 24/7 conversion facilities~\cite{circle2024buidl}, matches unlimited-liquidity Aave, with no failures.

At 10\% reserves, immediate settlement falls from 99.99\% with instant redemption to 93.0\% (T+0) and 85.1\% (T+1). The worst shock start times are Thursday evening to early Saturday (\autoref{fig:calendar}). Saturday-noon shocks cause 3.4\% and 5.4\% failures at 10\% reserves. Raising reserves to 30\% (T+0) or 40\% (T+1) keeps 10\%-shock failures at or below 0.03\% for all tested start times, but a 25\% shock still causes up to 2.4\% and 1.9\% failures (\autoref{tab:calendar}). The 24-hour demand-aware policy contains the 10\% shock under T+0. A 72-hour horizon avoids all failures but holds about 80\% of deposits to cover the settlement gap.

\begin{table}[H]
\centering
\begin{threeparttable}
\caption{Treasury-bill fund performance (\%).}
\label{tab:calendar}
\scriptsize
\setlength{\tabcolsep}{3pt}
\begin{tabular*}{\textwidth}{@{\extracolsep{\fill}}lcccccccc@{}}
\toprule
 & \multicolumn{2}{c}{\makecell{\textbf{Settled}\\\textbf{immediately}}} & \multicolumn{2}{c}{\textbf{Yield retained}} & \multicolumn{2}{c}{\makecell{\textbf{Failures}\\\textbf{10\% shock}}} & \multicolumn{2}{c}{\makecell{\textbf{Failures}\\\textbf{25\% shock}}} \\
\cmidrule(lr){2-3}\cmidrule(lr){4-5}\cmidrule(lr){6-7}\cmidrule(lr){8-9}
\textbf{Reserve policy} & T+0 & T+1 & T+0 & T+1 & T+0 & T+1 & T+0 & T+1 \\
\midrule
\textbf{Fixed 10\%} & 93.00 & 85.13 & 92.6 & 94.1 & 4.20 & 8.18 & 8.45 & 11.33 \\
\textbf{Fixed 20\%} & 99.80 & 99.32 & 82.3 & 83.7 & 0.84 & 4.59 & 5.08 & 7.82 \\
\textbf{Fixed 30\%} & 100.00 & 99.99 & 72.0 & 73.1 & 0 & 1.54 & 2.42 & 4.52 \\
\textbf{Fixed 40\%} & 100.00 & 100.00 & 61.5 & 62.5 & 0 & 0.03 & 0.30 & 1.87 \\
\textbf{Demand-aware, 24\,h} & 99.98 & 99.95 & 70.9 & 72.5 & 0 & 1.00 & 2.20 & 4.22 \\
\textbf{Demand-aware, 72\,h} & 100.00 & 100.00 & 20.8 & 22.3 & 0 & 0 & 0 & 0 \\
\bottomrule
\end{tabular*}
\begin{tablenotes}[flushleft]
\scriptsize
\item Settlement and retained yield: normal operation. Failures: worst case across shock start times.
\end{tablenotes}
\end{threeparttable}
\end{table}

\clearpage
\begin{figure}[H]
\centering
\includegraphics[width=0.82\textwidth]{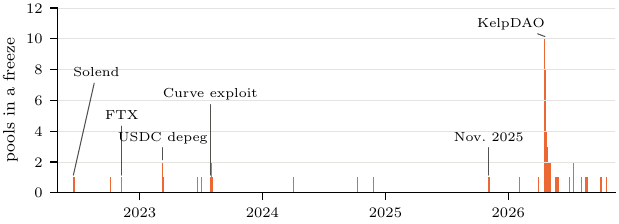}
\caption{Stablecoin lending-pool freezes.}
\label{fig:census}
\end{figure}

\subsubsection{Further Freezes.} We examine 64 DefiLlama stablecoin pools whose reported value equals available liquidity and whose October~2026 supply exceeds \$10~million: 17 chains, 18 protocols, and 45{,}469 pool-days since May~2022. A freeze requires liquidity below 5\% of its trailing 30-day median (at least \$5~million) and a supply rate at least 1.5 times its median. \autoref{fig:census} identifies 46 freezes in 27 pools, about one every two years per pool. Twenty-one last more than a day. April~2026 affected 14 pools on seven chains, up to ten on one day, limiting diversification within one protocol. The sample includes only surviving pools.

Using the RQ3 setup, we replay the Solend crisis of June~2022 and the Curve exploit of 30~July~2023, starting shocks at noon the day before each freeze and scaling to the pre-freeze pool size (\autoref{tab:freezes}). Solend's USDC liquidity fell from \$119~million to \$12{,}000 during a three-day freeze. At full capture neither design fails. At 10\% or 1\% capture, yield-token wallets fail on 7.4\% or 16.7\%, whereas AgentBank avoids failures with 20\% or 40\% reserves. The one-day Curve-related Aave freeze causes no failures under a 10\% shock. Under a 25\% shock at 1\% capture, AgentBank reserves below 10\% produce up to 1.8\% failures while yield-token wallets avoid them, because recalls occur every six hours.

\begin{table}[H]
\centering
\begin{threeparttable}
\caption{Payment failures under a 10\% withdrawal shock (\%).}
\label{tab:freezes}
\scriptsize
\begin{tabular*}{\textwidth}{@{\extracolsep{\fill}}llcccccc@{}}
\toprule
 & & \textbf{Yield-token} & \multicolumn{4}{c}{\textbf{AgentBank}, $r_{\mathrm{base}}$} & \textbf{Demand-} \\
\cmidrule(lr){4-7}
\textbf{Freeze} & \textbf{Capture} & \textbf{wallets} & \textbf{10\%} & \textbf{20\%} & \textbf{30\%} & \textbf{40\%} & \textbf{aware, 24\,h} \\
\midrule
\textbf{KelpDAO 2026}, Aave (5 days) & 100\% & 7.76 & 0.51 & 0 & 0 & 0 & 0 \\
 & 10\% & 12.83 & 4.56 & 1.68 & 0 & 0 & 0 \\
 & 1\% & 28.24 & 8.05 & 5.34 & 2.45 & 0.06 & 1.80 \\
\midrule
\textbf{Solend 2022}, Solana (3 days) & 100\% & 0 & 0 & 0 & 0 & 0 & 0 \\
 & 10\% & 7.38 & 1.24 & 0 & 0 & 0 & 0 \\
 & 1\% & 16.72 & 4.66 & 1.99 & 0.10 & 0 & 0.28 \\
\midrule
\textbf{Curve exploit 2023}, Aave (1 day) & 1--100\% & 0 & 0 & 0 & 0 & 0 & 0 \\
\bottomrule
\end{tabular*}
\begin{tablenotes}[flushleft]
\scriptsize
\item Aave: USDC on Ethereum. Solana: Solend USDC. Capture: share of available pool liquidity.
\end{tablenotes}
\end{threeparttable}
\end{table}

\clearpage
\subsubsection{Priority-Fee Race.} When automated agents react in the same block, priority fees replace reaction delay in determining order~\cite{daian2020flashboys}. We model a complete-information first-price auction among the run-study depositors (mean balance \$2{,}000). The largest unserved balance sets the price paid by every fully served automated depositor. A race begins when automated claims exceed assets, for example at 95\% automated balances and $\ell=5\%$. With all balances automated, prices are \$284, \$526, and \$900 at losses of 2\%, 5\%, and 10\%. Fees consume 10.1\%, 14.7\%, and 18.6\% of deposits, or 1.9--5.0 times the loss, reducing aggregate recovery to 71--88\% from 90--98\%. The top balance decile recovers 92--98\% net of fees, while 46--75\% of automated depositors lose over half their balance. Complete information makes these fees an upper bound. An epoch spanning the automated reaction window removes ordering value and restores $A/L$ recovery. Solvency-contingent redemption also removes the race.

\begin{figure}[H]
\centering
\includegraphics{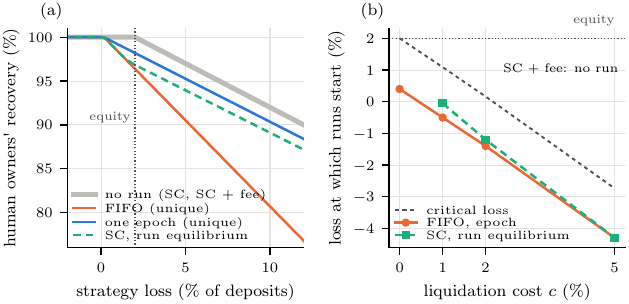}
\caption{Runs under noisy signals: \textbf{(a)}~human-owner recovery; \textbf{(b)}~run thresholds.}
\label{fig:global}
\end{figure}

\subsubsection{Noisy Signals.} Common knowledge of losses permits multiple run equilibria~\cite{diamond1983bankRuns}. Following global-games models~\cite{goldstein2005demanddeposit,rochet2004coordination}, depositors instead observe $x=\ell+\eta\varepsilon$, where $\varepsilon$ is standard normal, under an uninformative prior. They redeem when expected payout, net of a 0.05\% balance cost, exceeds recovery from waiting. Equity is $e=2\%$, reserves $r=10\%$, and half the balances are automated and served first. Payouts beyond reserves liquidate assets at discount $c$, creating strategic complementarity~\cite{chen2010payoff}. We iterate best responses from ``nobody redeems'' and ``everybody redeems''. Coincident limits identify a unique threshold equilibrium.

\autoref{fig:global} uses $\eta=0.5\%$. Noise of 0.25--2\% and liquidation discounts up to 5\% give the same pattern. FIFO iterations coincide within grid resolution, with runs starting 3.0--3.3 signal standard deviations below the insolvency threshold $e-(1-r)c/(1-c)$. At $c=2\%$, that threshold is near zero. A 3\% loss leaves humans 94.4\% rather than 99\% recovery. One epoch leaves the threshold unchanged but halves human losses. Solvency-contingent redemption has a unique no-run equilibrium if liquidation is costless. For $c>0$, early redeemers avoid costs borne by later claims, allowing a run even without losses. The liquidity fee removes the run for all tested $c$ and $\eta$.

\begin{figure}[H]
\centering
\includegraphics{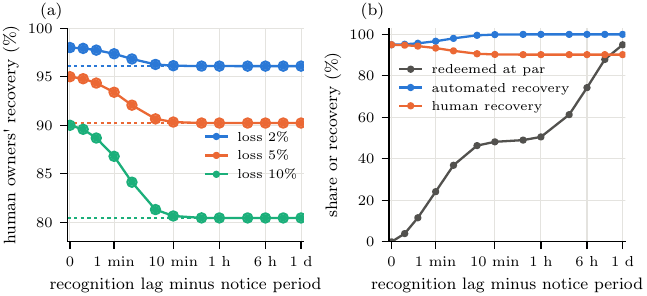}
\caption{Recognition lag: \textbf{(a)}~human recovery; \textbf{(b)}~par redemptions and recovery.}
\label{fig:lag}
\end{figure}

\subsubsection{Recognition Lag.} We delay reported losses after an impairment becomes known off-chain, using the RQ4 reaction delays. All assets are liquid. Reserves are 10\%, and haircut factors of 0.98, 0.95, and 0.90 require equity of 2\%, 4.7\%, and 10\% of deposits. As \autoref{prop:runs}(iii) predicts, early settlements remain at par. For a 5\% loss, a one-minute lag lowers human recovery from 95.0\% to 93.4\% (\autoref{fig:lag}). At 30~minutes, all automated agents have redeemed at par and human recovery falls to 90.2\%, matching FIFO. With longer lags, humans also redeem at par and withdrawals reach 74\% of balances after six hours. Outcomes are identical across haircuts. Forward pricing delays every settlement by the notice period, shortening the effective recognition lag. A notice period longer than the lag restores solvency-contingent settlement.

\end{document}